\documentclass[aps,prl,twocolumn,superscriptaddress]{revtex4}
\usepackage{graphicx}
\usepackage{dcolumn}
\usepackage{bm}
\usepackage{color}
\usepackage{hyperref}
\usepackage{amsmath}
\usepackage{amssymb}

\begin{document}

\title{
Laminar, turbulent and inertial shear-thickening regimes in channel flow \\ of neutrally buoyant particle suspensions }

\author{Iman Lashgari}\email[]{imanl@mech.kth.se}
\affiliation{Linn{\'e} FLOW Centre and SeRC, KTH Mechanics, Stockholm, Sweden}
\author{Francesco Picano}
\affiliation{Linn{\'e} FLOW Centre and SeRC, KTH Mechanics, Stockholm, Sweden}
\affiliation{Department of Physics, Sapienza University of Rome, Italy}
\author{Wim-Paul Breugem}
\affiliation{Laboratory for Aero \& Hydrodynamics, TU-Delft, Delft, The Netherlands}
\author{Luca Brandt}
\affiliation{Linn{\'e} FLOW Centre and SeRC, KTH Mechanics, Stockholm, Sweden}

\date{\today}

\begin{abstract}

The aim of this Letter is to characterize the flow regimes of suspensions of finite-size {rigid} particles in a viscous fluid at finite inertia.
We explore the system behavior as function of the particle volume fraction and the Reynolds number (the ratio of flow and particle inertia to viscous forces). Unlike single phase flows where a clear distinction exists between the laminar and the turbulent states, three different {regimes} can be identified in the presence of a particulate phase,  with smooth transitions between them. At low volume fractions, the flow becomes turbulent when increasing the Reynolds number, transitioning from the laminar regime dominated by viscous forces to the turbulent regime characterized by enhanced momentum transport by turbulent eddies. At larger volume fractions, we identify a new {regime} characterized by an even larger increase of the wall friction. The wall friction increases with the Reynolds number (inertial effects) while the turbulent transport is weakly affected, as in a state of intense inertial shear-thickening. {This state may  prevent the transition to a fully turbulent regime at arbitrary high speed of the flow.}
\end{abstract}


\maketitle

Understanding transition from laminar to turbulent flow has puzzled many scientists since the seminal work by Reynolds \cite{Reynolds83}.
Despite the vast number of investigations, this phenomenon 
is not well understood even in the simplest configurations as plane 
channel flow.    
To be able to predict, and possibly control, the onset of turbulence is crucial in numerous applications as this is associated with a sharp increase of the wall friction and of the total drag.
Transition in channel flows is of subcritical nature, 
 it occurs if the flow is
forced by strong enough perturbations,  and takes place at Reynolds numbers, $Re$,
significantly lower than that predicted by linear stability \cite{Eckhardt07} ($Re$ quantifies the ratio of inertial to viscous forces).

When a suspension of rigid particles is considered instead of a pure fluid, 
{the particle-fluid interactions significantly alter the bulk behavior of the suspension \cite{mor_ra09}  and} 
peculiar and unexplained effects appear in the
transitional regime. These aspects, important in several environmental and industrial applications, 
will be investigated in  this Letter.

Matas et al. \cite{Matas03} performed experiments on the laminar-turbulent transition in particle-laden pipe flows. For 
sufficiently large particles, they found  
a non-monotonic variation of the transitional Reynolds number when increasing the particle volume fraction $\Phi$ (the bulk volume fraction). 
They were not able to collapse the transitional Reynolds numbers to the one 
of the single-phase fluid by rescaling with the suspension viscosity for the larger particles used, and thus suggested the presence of some additional dissipation mechanisms, still to be explored.
They also underlined the difficulties found to unambiguously define the flow regime 
(laminar or turbulent) since suspensions manifest large fluctuations also at low speeds.

Similar findings were recently reproduced numerically in \cite{Yu13}.
These authors also note how arbitrary it is to {define laminar and turbulent regimes} in suspensions because the value of the transitional Reynolds number  is very sensitive to the velocity fluctuation threshold chosen to discern the flow state. 
The characteristics of the different flow regimes in particle-laden flows and the mechanisms behind the transition are  
not yet understood.

In this Letter we show the existence of three different flow {regimes} in the presence of a particulate phase with smooth transitions between them. First, a laminar-like regime (low $Re$ and $\Phi$), dominated by viscous forces, and, second, a turbulent-like regime (high $Re$ and 
sufficiently low $\Phi$), characterized by increased wall friction due to the turbulent transport, classically quantified by the Reynolds stresses \cite{Pope00}. The third state appears in the dense regimes (high $\Phi$), where we observe  
a significant increase of the wall friction, larger than in the turbulent-like regime. This is not associated with the increase of the Reynolds stresses, but to an increase of the particle-induced stress indicating a transport mechanism different than that of turbulence. Following rheology literature, we recognize this state as an intense 
\emph{inertial shear-thickening}: it corresponds to an increase of the wall friction (effective viscosity) 
with the flow shear rate (Reynolds number) due to increasing particle stresses with inertia; thus the additional dissipation mechanism suggested in \cite{Matas03}.  
 
The analysis is based on data from fully resolved direct numerical simulations of a channel flow laden with 
rigid {neutrally-buoyant} spherical particles.
The code used, the Immersed Boundary solver developed by Breugem \cite{Breugem12},  couples an uniform Eulerian mesh for the fluid phase with a Lagrangian mesh for the solid phase.
Lubrication corrections and a soft-sphere collision model have been implemented to address interactions among the particles when their relative gap distance is below the mesh size. We model a surface roughness layer around the particle whose thickness is $10^{-2} $ the particle radius and where the lubrication correction is kept fixed. This is turned off when the collision starts 
to occur, see~\cite{Lambert13} for more details. The code has been validated against several test cases over a range of Reynolds numbers \cite{Breugem12} and recently used to study  the rheology of dense and active suspensions \cite{Picano13,Lambert13}.

We simulate a channel flow with periodic boundary conditions  in the streamwise and spanwise directions.  
Here, the streamwise, wall-normal and spanwise coordinates and velocities are denoted by  $y,z,x$ and $v,w,u$, respectively.  
The size of the computational domain is $6h \times 2h \times 3h $ where $h$ is the half-channel {height}. The diameter of the particles is set to a fixed value, $2a= 2h/10$ matching the experiment with largest particles in \cite{Matas03}. The number of uniform Eulerian grid points 
 is $480 \times 160 \times 240$ in the streamwise, wall-normal and  spanwise directions; 746 Lagrangian grid points have been used on the surface of each particle to resolve the interactions between the two phases. We define the Reynolds number by the bulk velocity $U_b$, channel height and the fluid kinematic viscosity $\nu$: $Re=2U_b h/\nu$.  
The dataset explores a wide range of Reynolds numbers,
$500\le Re \le5000$, and particle volume fractions $0\le \Phi\le0.3$.

The initial condition for the simulations is chosen to be a high-amplitude (maximum normal velocity equals to $U_b$) localized disturbance in the form of pair of streamwise vortices  \cite{Henningson91}. The critical value for transition of the single phase flow in the present setup is found to be $Re\approx2300$; the velocity fluctuations vanish in time for Reynolds number $Re<2300$ while they reach a finite amplitude for $Re> 2300$.

\begin{figure}[t]
\includegraphics[width=0.45\linewidth]{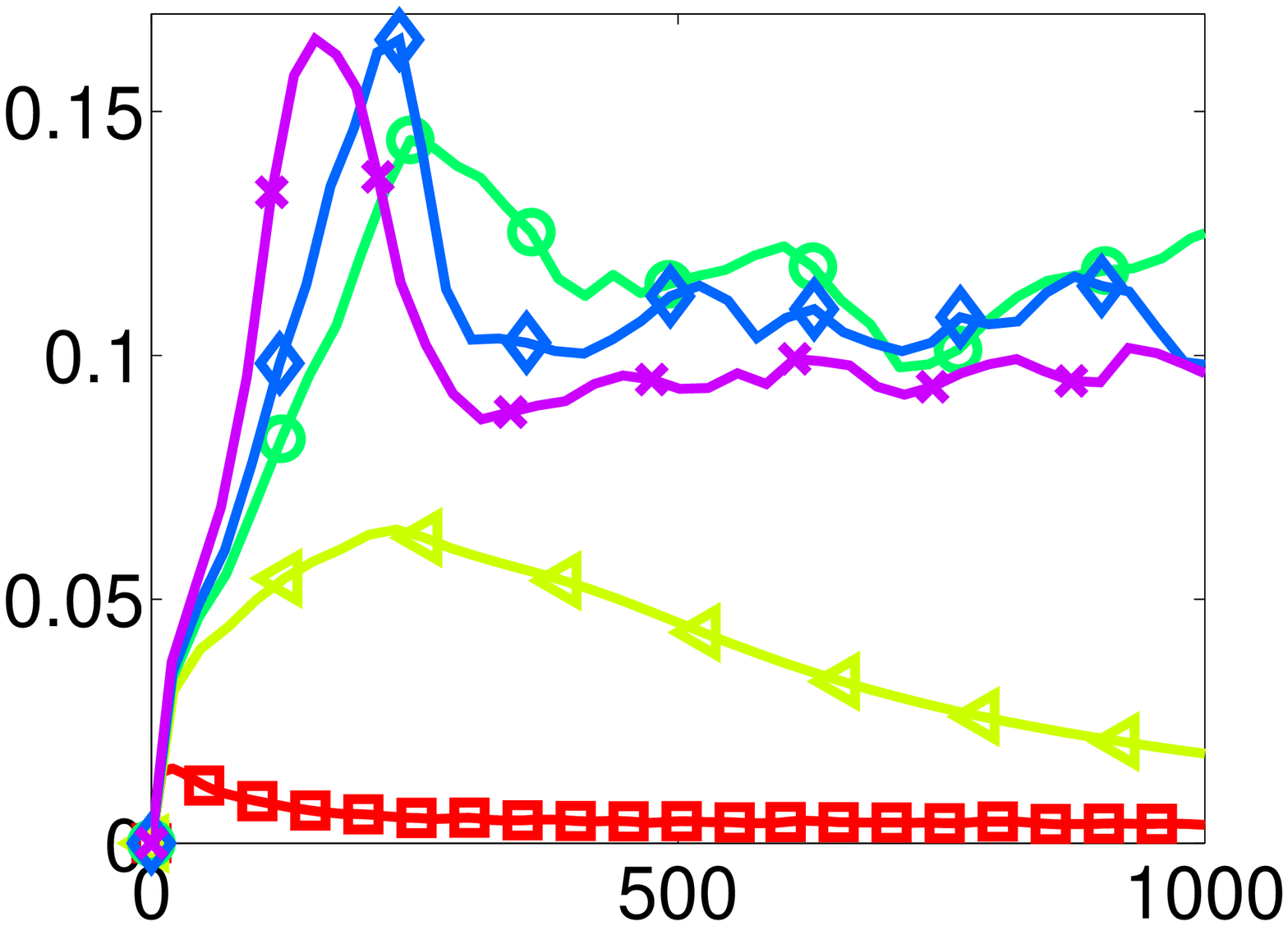}
\put(-110,80){{\large $(a)$}}
\put(-125,40){{$\tilde v_{rms}$}}
\put(-50,-10){{$t$}}
\includegraphics[width=0.45\linewidth]{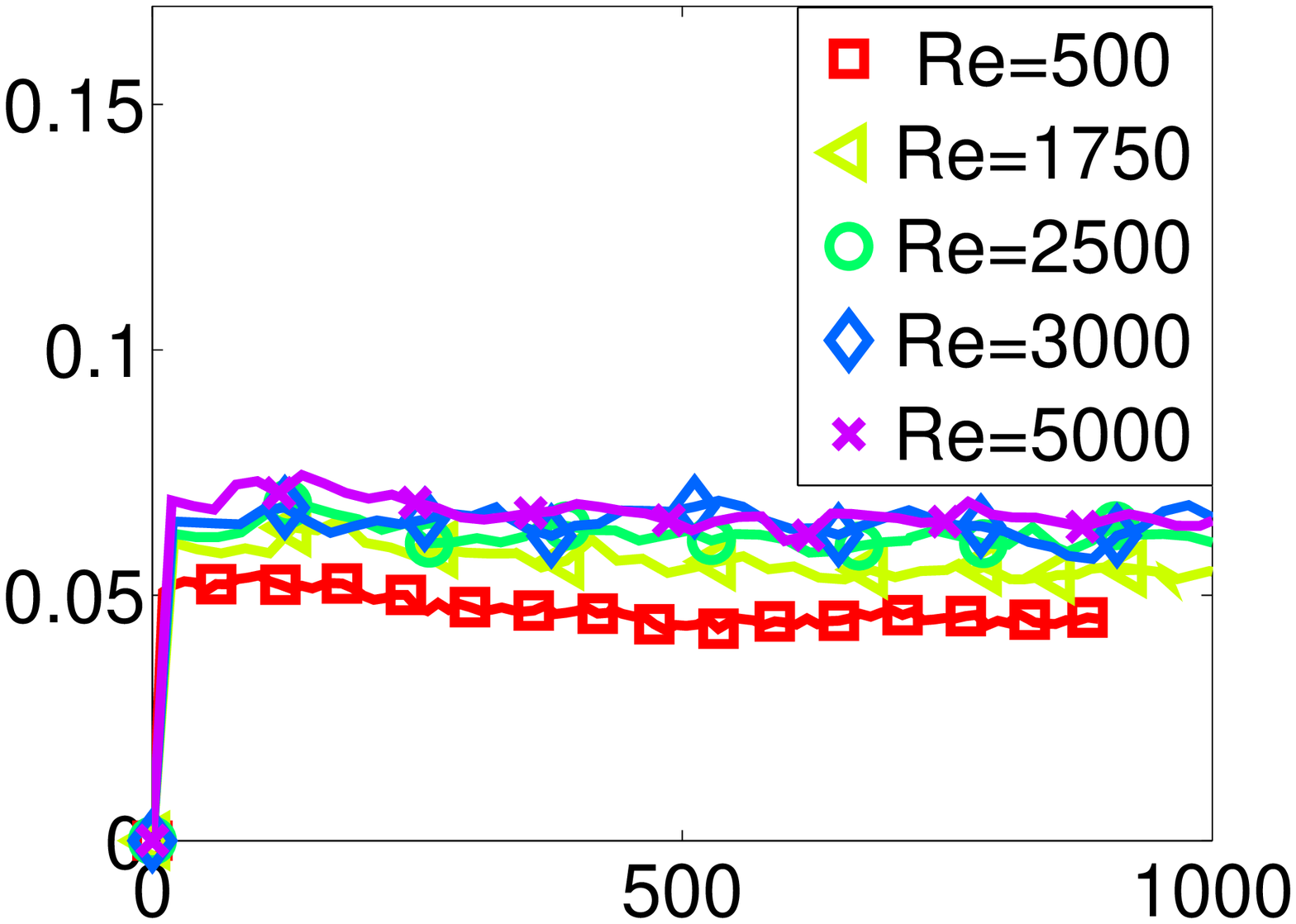}
\put(-110,80){{\large $(b)$}}
\put(-50,-10){{$t$}}
\caption{\label{fig:rms} 
Time series of the streamwise velocity fluctuation $\tilde v_{rms}$ for (a) $\Phi=0.001$ and (b) $\Phi=0.3$.}
\end{figure}

To document the difficulty of clearly defining a transition threshold at high volume fractions,
time histories of the root-mean-square of the streamwise velocity fluctuations, $\tilde v_{rms}(t)$({box-averaged}), 
are displayed in figure~\ref{fig:rms} for low and high volume fractions (velocity fluctuations are normalized by $U_b$ and time by $h/U_b$).
For $\Phi=0.001$, the transition to turbulence follows classic results in single phase fluids. For $Re\geq2000$ the time histories exhibit a transient peak after which the level of fluctuation saturates to the values of the turbulent regime, $0.1<\tilde v_{rms}<0.12$, just weakly dependent of the Reynolds number. 
Interestingly, the presence of few finite-size particles in the flow create continuous background disturbances such that the threshold at which turbulence can be sustained decreases 
to $1750<Re<2000$. 
Figure~\ref{fig:rms}(b) shows that the level of streamwise velocity fluctuations saturate to $0.05<\tilde v_{rms}<0.07$ for all the Reynolds numbers under investigation when $\Phi=0.3$. 
The velocity fluctuations in the flow smoothly increase with the Reynolds number and approach a regime value lower than that of the turbulent flow at small $\Phi$.  As shown in the plot and discussed recently in \cite{Yu13}, the definition of a transitional Reynolds number becomes very sensitive to the threshold values chosen, and probably not completely meaningful as we show below.

\begin{figure}[t]
\includegraphics[width=1.0\linewidth]{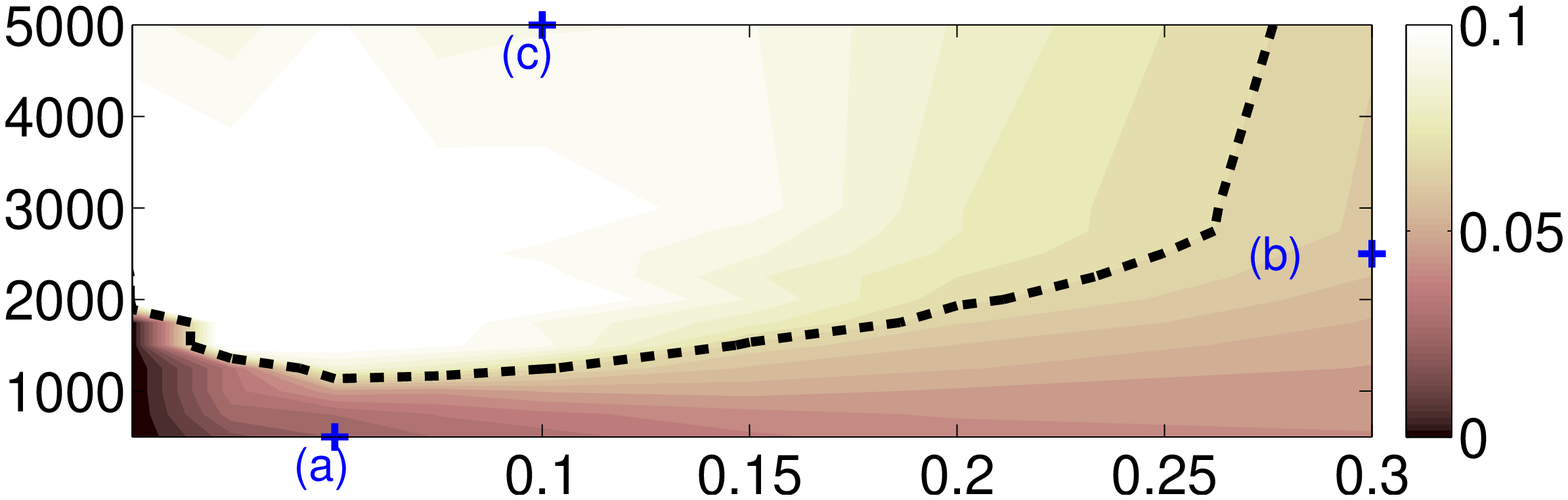}
\put(-250,70){{\large $(a)$}}
\put(-245,35){{$Re$}}
\\
\includegraphics[width=1.0\linewidth]{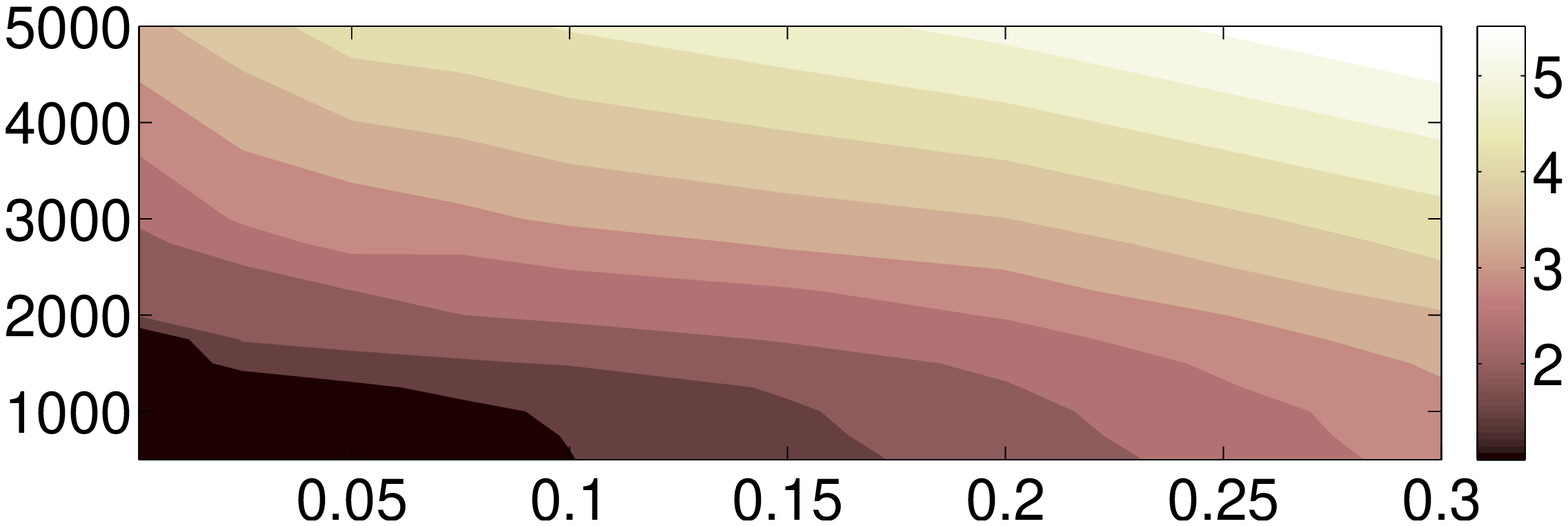}
\put(-250,68){{\large $(b)$}}
\put(-245,35){{$Re$}}
\put(-130,-10){{$\Phi$}}
\caption{\label{fig:effective-viscosity} 
 (a) Contour map of the streamwise velocity fluctuations $v_{rms}$ and (b) Relative viscosity $\nu_r$ in the $(Re,\Phi)$ plane.{ In total, 96 points have been simulated in the $(Re,\Phi)$ plane.}}
\end{figure}

A contour map of the time-averaged streamwise velocity fluctuation intensity $v_{rms}=\langle v'^2 \rangle$ in the $(Re,\Phi)$ plane is presented in figure~\ref{fig:effective-viscosity}(a). {The fluctuation intensity is based on the mixture of the fluid and particle velocities and obtained by a classical ensemble average, $\langle \, \rangle$.} 
The samples for the ensemble average are collected after the transient phase of the disturbance evolution. 
The black dashed-line {on the map} represents the threshold value 0.07 that can be reasonably used to identify the chaotic regime.
This shows a non-monotonic behavior of the transitional Reynolds number, in line with the experimental observation in \cite{Matas03} and the simulations in \cite{Yu13}. 
We have marked three points on the map corresponding to the (a) laminar-like, (b) inertial shear-thickening and (c) turbulent-like regime for future reference. 
Figure~\ref{fig:effective-viscosity}(b) displays the average wall friction $\tau_w(Re,\Phi)$ normalized by the friction of the laminar single-phase  Poiseuille  flow $\tau^0_w(Re)$. 
We thus define a relative viscosity 
 $\nu_r=\tau_w/\tau^0_w$, similarly to what is usually adopted in rheology~\cite{stipow_arfm05}.
A jump in the value of the relative viscosity, to about the double, occurs at the critical conditions at low volume fractions {(e.g. $\Phi=0.05$)}. Increasing the particle concentration, the relative viscosity smoothly increases with the Reynolds number (inertial effects) while the level of fluctuations, yet significant, remains almost unaffected. This behavior points toward a new dynamic state that we call \emph{intense inertial shear-thickening}. 
It is important to note that 
 the values of $\nu_r$ for $Re=500$ ({laminar flow,} weak inertia) well follow classic semi-empirical fits  
for viscous laminar suspensions, e.g.\ Eilers fit~\cite{stipow_arfm05}, when $\Phi \leq 0.2$. For $\Phi=0.3$, the comparison with the data of Yeo et. al. \cite{yeomax_jfm11} reveals some shear thickening.

To characterize the different regimes of the particulate flow, we study the stress budget in the flow. Based on the formulation proposed in  \cite{Zhang10}, inspired from the early work by Batchelor\cite{Batchelor70}, we write the momentum balance in the wall-normal direction assuming streamwise and spanwise homogeneity,
\begin{align}\label{equ:stress}
\tau(z/h)= - (1-\varphi)\langle w'^{f}v'^{f}\rangle - \varphi\langle w'^{p}v'^{p}\rangle  +
\\ \nonumber
 \nu(1-\varphi)\frac{\partial V^f}{\partial z} + \frac{\varphi}{\rho}\langle \sigma_{yz}^p\rangle = \nu \left.\frac{\partial V^f}{\partial z}\right|_w \left(1-\frac{z}{h}\right),
\end{align}
where $\varphi(z)$ denotes the mean local volume fraction, the superscripts $f$ and $p$  the fluid and particle phase, $\rho$ {the fluid mass density }and 
$\sigma_{yz}$ the particle stress. The first two terms represent the contribution of both phases to the Reynolds shear 
stress: $\tau_R(z)$. The third and fourth term are the viscous, $\tau_V(z)$, and particle stresses, $\tau_P(z)$. 
The particle stress contains contributions from the hydrodynamic stresslet, particle acceleration and inter-particle stresses (see\cite{Zhang10,Batchelor70}). 
The total stress is balanced by the external streamwise pressure gradient, yielding a linear total stress profile across the channel with maximum at the wall, 
$\tau_w=\nu \left.\frac{\partial V^f}{\partial z}\right|_w$, and zero at the centerline.

\begin{figure}[t]
\includegraphics[width=1.0\linewidth]{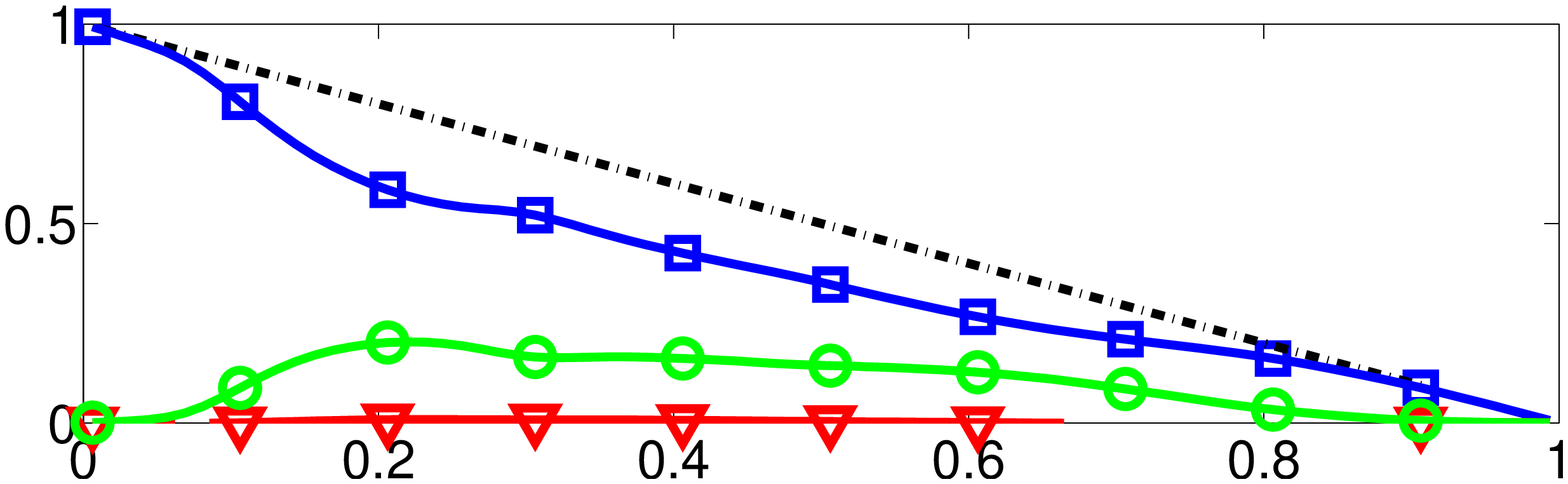}
\put(-235,60){{\large $(a)$}}
\\
\includegraphics[width=1.0\linewidth]{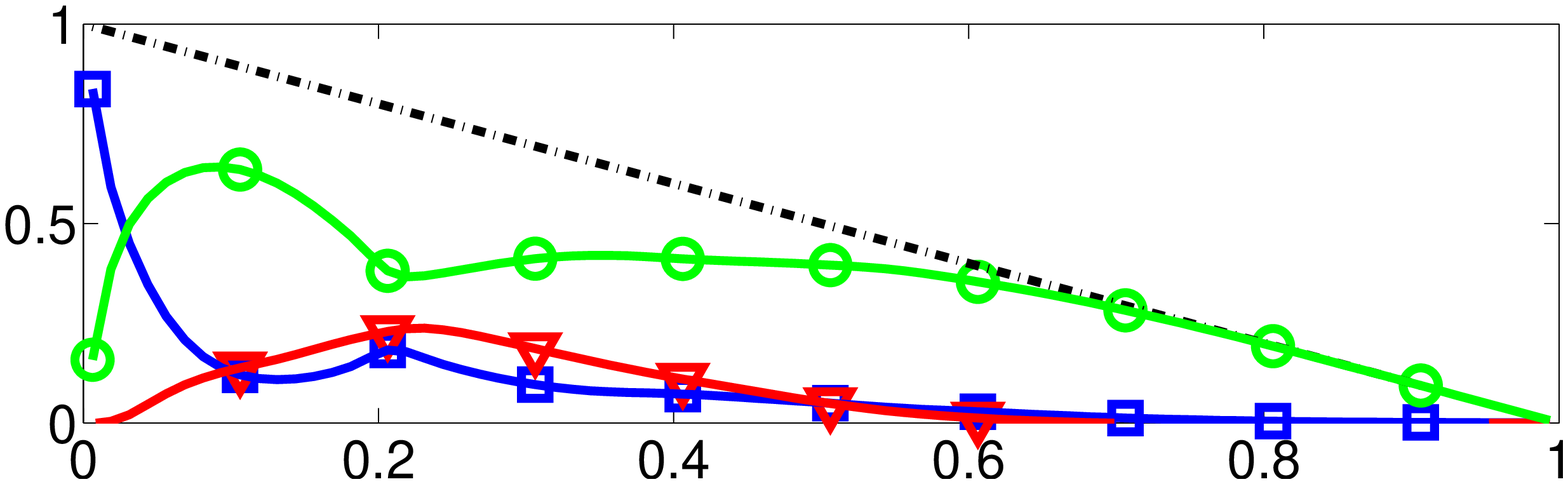}
\put(-235,60){{\large $(b)$}}
\\
\includegraphics[width=1.0\linewidth]{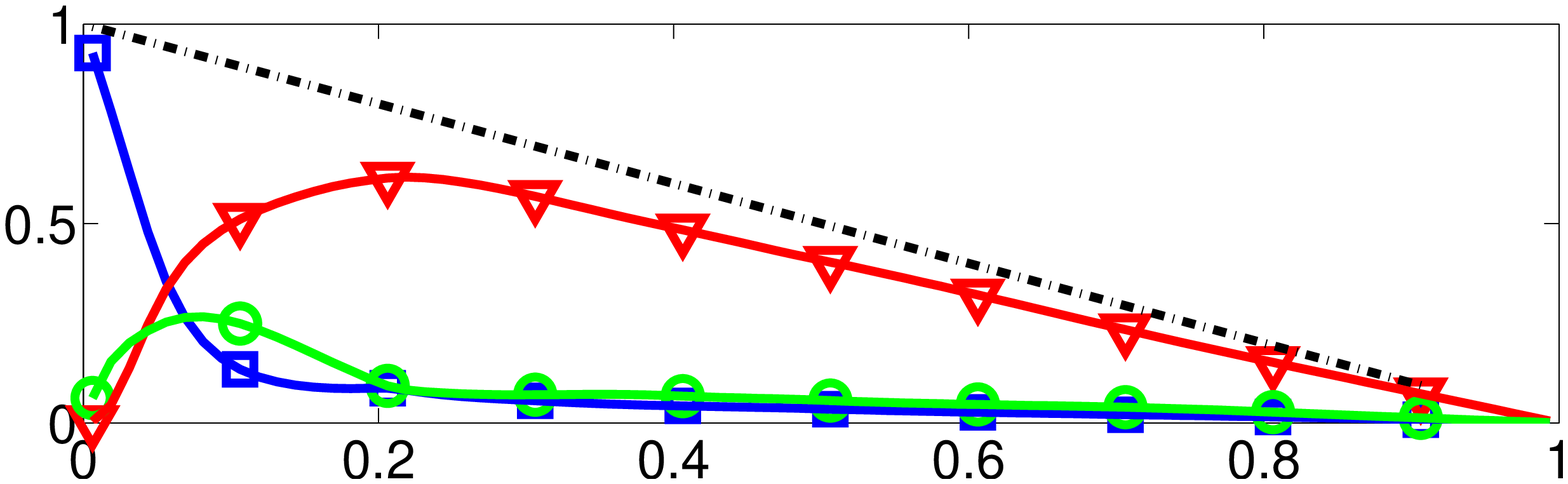}
\put(-235,60){{\large $(c)$}}
\put(-130,-10){{$z/h$}}
\\
\caption{\label{fig:stress-budget} 
Budget of the momentum transport for (a) $Re=500$ \& $\Phi=0.05$ (b) $Re=2500$ \& $\Phi=0.3$ (c) $Re=5000$ \& $\Phi=0.1$, (red $\triangledown$): Reynolds stress, (blue $\square$): viscous stress and (green $\circ$): particle stress. }
\end{figure}

We evaluate the contribution of each term in the stress budget for the cases marked by a,b and c in figure~\ref{fig:effective-viscosity}(a) and display 
the wall-normal stress profiles, normalized by $\tau_w$, in fig.~\ref{fig:stress-budget}. For the laminar-like regime,  the Reynolds shear stress is 
negligible and the viscous stress dominates (see panel a).
Figure~\ref{fig:stress-budget}(c) displays the stress budget of the turbulent-like regime when the Reynolds stress is the dominant term and both 
viscous and particle stresses are relevant only in the near-wall region. This is the typical behavior of turbulent flows~\cite{Pope00}:
the increase of the wall friction $\tau_w$ with respect to a laminar state is 
caused by the formation of coherent velocity fluctuations and increased streamwise momentum transport in the wall-normal direction.     
For the inertial shear-thickening regime, figure~\ref{fig:stress-budget}(b), the particle stress is larger than the Reynolds shear stress, and accounts for 
the majority of the momentum transfer (78\%) when integrated across the channel. In the middle of the channel, $z \ge 0.6h$, the 
particle stress is the only term to transfer stress exceeding by more than one order of magnitude the viscous and the Reynolds stresses. This is 
due to the particle migration that leads to a mean local volume fraction $\varphi\simeq0.5$ 
at the channel centre 
(not shown here), see \cite[][]{brojae_prl09,Fall10}.
 {  High particle concentration near the wall induces strong particle-wall interactions and a peak in the profile of particle stress. In the mid region between the wall and the centreline, the turbulent stress is largest: the contribution of the particle Reynolds stress is much smaller than that of the fluid Reynolds stress (not shown) suggesting that the momentum is transferred by the fluid and the coherent motion of the particles is less important.  }

\begin{figure}[t]
\includegraphics[width=1.0\linewidth]{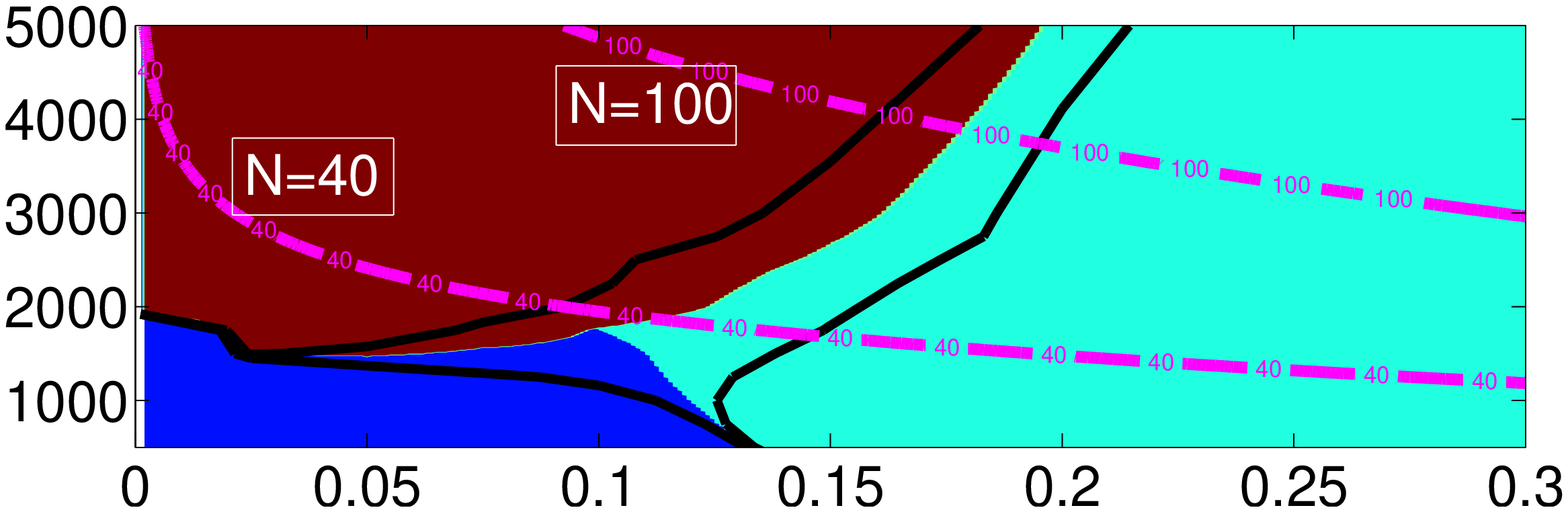}
\put(-250,65){{\large $(a)$}}
\put(-245,35){{$Re$}}
\put(-120,-10){{$\Phi$}}
\\
\vspace{.2cm}
\hspace{-1cm}
\includegraphics[trim = 5mm 30mm 30mm 10mm, width=0.8\linewidth]{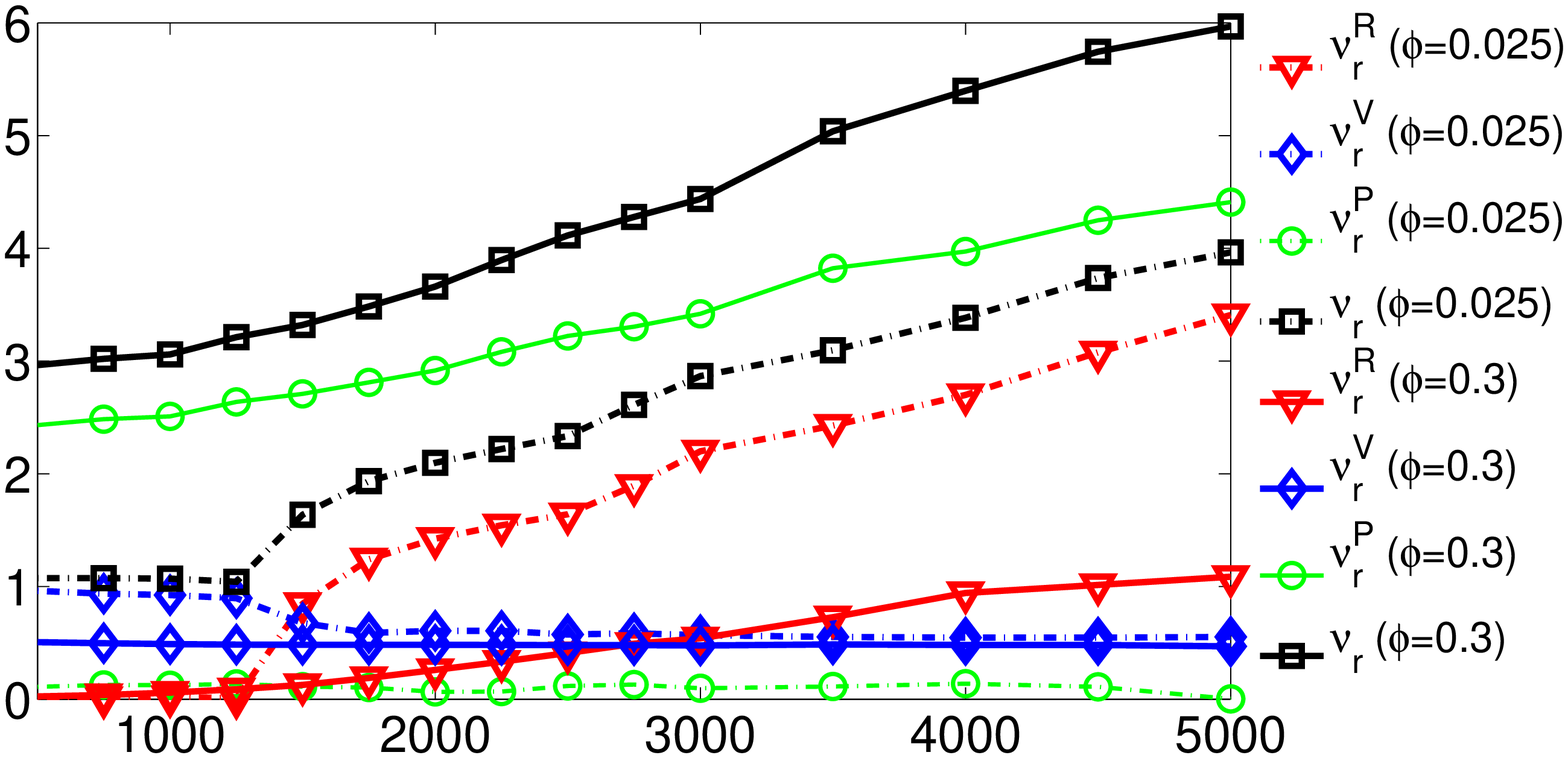}
\put(-210,55){{\large $(b)$}}
\put(-190,20){{$\nu_r$}}
\put(-100,-30){{$Re$}}
\\
\caption{\label{fig:phase} 
(a) Phase diagram for particle laden channel flow 
in the $Re-\Phi$ plane (blue, red and green colors indicate laminar-like, turbulent-like and inertial shear-thickening regime). The solid black lines show the boundary of the regions where each term in the stress budget is over $50\%$ of the total stress. The dashed purple lines are iso-levels of Bagnold number. 
(b) Contributions to the total wall friction from the viscous stresses, Reynolds stresses and particle stresses versus the Reynolds number $Re$ for $\Phi=0.025$ and $\Phi=0.3$. }
\end{figure}

Using the momentum transfer budget we are therefore in the position to suggest a
quantitative
classification of the flow. 
In particular, dividing equation~\eqref{equ:stress} with $\tau=\tau_w\, (1-z/h)$ and integrating across the channel
\begin{equation}
1= \overline{\tau_V/\tau}+\overline{\tau_R/\tau}+\overline{\tau_P/\tau}=\Sigma_V+\Sigma_R+\Sigma_P.
\label{eq:mean} 
\end{equation}
The flow is denoted as laminar-like if the viscous stress gives the largest contribution to the total momentum transfer
across the channel (relative majority), i.e.\ $\Sigma_V> \Sigma_R$ and $\Sigma_V>\Sigma_P$,
and similarly, turbulent-like if the Reynolds shear stress provides the largest contribution. We denote as 
intense inertial shear-thickening the regime where the momentum transfer is dominated by the particle stress, $\Sigma_P>\Sigma_V$ and $\Sigma_P>\Sigma_R$.  
The region of existence of these three regimes is depicted in the $(Re,\Phi)$  plane in figure~\ref{fig:phase}(a). 
The solid black lines show the boundary of the regions where each term in the stress budget is over $50\%$ of the total 
(absolute majority).
Note that the phase diagram is computed for $a/h=1/10$ and the exact values of $\Phi$ at which transitions occur are dependent on this ratio.

At vanishing volume fractions, $\Phi\rightarrow0$, 
the transition between the laminar and turbulent regime is, as expected, sharp. Indeed, criteria based either on fluctuation levels or on the stress budget determine the same critical Reynolds number. At finite $\Phi$, the transition among the states becomes smooth. In this case, the definition of the transitional Reynolds number would depend on the threshold value chosen for the velocity fluctuations and on the choice of the velocity component. In this scenario, the
stress balance can quantitatively identify the dominant interactions.

The flows determined by the particle stress corresponds to an intense form of continuous shear-thickening. Inertial shear-thickening amounts to an increase of the relative viscosity $\nu_r=\tau_w/\tau^0_w$ with 
the imposed shear rate $\dot \gamma$ (Reynolds number) at fixed $\Phi$ without an increase of the Reynolds stresses (turbulent activity)~
\cite{stipow_arfm05,Picano13}. 
Note that the shear rate $\dot \gamma=U_b/h$ is proportional to the Reynolds number $Re=2\dot \gamma h^2/\nu$.
Defining the relative viscosity as $\nu_r=\tau_w/\tau^0_w$, we can extract the contributions to the total wall friction due to 
the viscous $\nu_r^V=\Sigma_V\,\nu_r$, particle $\nu_r^P=\Sigma_P\,\nu_r$ and Reynolds stresses $\nu_r^R=\Sigma_R\,\nu_r$. 
These quantities are depicted vs $Re$ in figure~\ref{fig:phase}(b) for  $\Phi=0.025$ 
and $\Phi=0.3$. At low $\Phi$, the relative viscosity $\nu_r$ is constant up to $Re \approx 1500$ and determined 
mainly by the viscous contribution; its value is slightly higher than unity 
and matches the values observed in inertialess Couette flows. The shear-thickening one would expect to see in Couette flows at these  finite Reynolds number is canceled by a non-constant shear rate across the channel that can induce particle migrations. 
The increase of wall friction at higher $Re$ is due to the action of the Reynolds stresses and to the onset of a turbulent flow.

The situation dramatically changes at $\Phi=0.3$ where we observe a smooth and continuous
increase of $\nu_r$ that can be closely correlated with the increase of the particle-induced stress, and only in minor part to the increase of
the Reynolds stress contribution. 
At sufficiently high $\Phi$, where $\Sigma_P$ dominates, an intense shear-thickening occurs causing a large increase of the wall
friction. 
{The data at $\Phi=0.3$ also reveal that the turbulent contribution to the total momentum transfer is appreciable and increases at a similar rate as the particle-induced contribution. This behavior suggests that in dense cases the inertial shear-thickening may  prevent the transition to a fully turbulent regime at arbitrary high speed of the flow.} 
This finding partially answers the 
conjecture left by Bagnold \cite{bagnold}: \textit{``It seems that the residual fluid shear stress due to turbulence progressively gives place to grain shear stress''} when increasing $\phi$ and $Re$.

The Bagnoldian regime is dominated by grain stress and occurs when the Bagnold number, $N= 2 Re (a/h)^{1/2}\lambda^{1/2}$ with $\lambda=1/[(0.74/\Phi)^{1/3}-1]$, is above
 450; when $N\le40$ the macro-viscous regime takes place. In our simulations, 
 the Bagnold number is higher than 40 and below 170 for most of the  turbulent and the particle dominated cases. The simulations denoted by b) and c) share almost the same Bagnold number $N\sim100$ and relative viscosity (see figure~\ref{fig:effective-viscosity}b and \ref{fig:phase}a), however the momentum transport and the underlying physics are completely different.
 The Bagnold number $N$ alone
 is therefore not sufficient to discriminate between turbulent and particle dominated regimes.

We note that inertial shear-thickening takes place only if inertial effects are present
at the particle size~\cite{stipow_arfm05,kulmor_pof08}, i.e.\ finite particle Reynolds number ($Re_a=\dot \gamma a^2/\nu$). The same behavior is not observed for small particles (yet large enough to be non-Brownian) when inertial effects are negligible ($Re_a \propto a^2$) and the flow around the particles remains in the Stokes regime. In this case the main effect of the solid phase on the flow can be well described by the
mixture effective viscosity as demonstrated in Matas et al.~\cite{Matas03} 
who scale the experimental  data for small particles only with the suspension viscosity measured in the laminar regime.

In summary, we study semi-dilute particulate flows and observe an increase of the normalized wall friction (relative viscosity) with increasing inertial effects not associated with a corresponding
increase of the Reynolds stresses. We identify this regime as an intense form of continuous inertial shear thickening, induced by inertial effects at the particle size at large enough nominal volume fractions, see e.g.~\cite{Picano13}. 
This leads to a phase diagram, function of $\Phi$ and $Re$, with three different regimes. 
In the limit of  vanishing volume fraction, the transition between laminar and turbulent regime is sharp, 
while this is not the case 
at finite $\Phi$. This implies that at moderate and high $\Phi$ inertial
shear-thickening and coherent turbulence coexist with different relevance.\\

This work was supported by the European Research Council Grant No.\ ERC-2013-CoG-616186, TRITOS and by the Swedish Research Council (VR). Computer time provided by SNIC (Swedish National Infrastructure for Computing) is gratefully acknowledged.


\begin{thebibliography}{18}
\expandafter\ifx\csname natexlab\endcsname\relax\def\natexlab#1{#1}\fi
\expandafter\ifx\csname bibnamefont\endcsname\relax
  \def\bibnamefont#1{#1}\fi
\expandafter\ifx\csname bibfnamefont\endcsname\relax
  \def\bibfnamefont#1{#1}\fi
\expandafter\ifx\csname citenamefont\endcsname\relax
  \def\citenamefont#1{#1}\fi
\expandafter\ifx\csname url\endcsname\relax
  \def\url#1{\texttt{#1}}\fi
\expandafter\ifx\csname urlprefix\endcsname\relax\def\urlprefix{URL }\fi
\providecommand{\bibinfo}[2]{#2}
\providecommand{\eprint}[2][]{\url{#2}}

\bibitem[{\citenamefont{Reynolds}(1983)}]{Reynolds83}
\bibinfo{author}{\bibfnamefont{O.}~\bibnamefont{Reynolds}},
  \bibinfo{journal}{Philos. Trans. R. Soc} \textbf{\bibinfo{volume}{174}}
  (\bibinfo{year}{1983}).

\bibitem[{\citenamefont{Eckhardt et~al.}(2007)\citenamefont{Eckhardt,
  Schneider, Hof, and Westerweel}}]{Eckhardt07}
\bibinfo{author}{\bibfnamefont{B.}~\bibnamefont{Eckhardt}},
  \bibinfo{author}{\bibfnamefont{T.}~\bibnamefont{Schneider}},
  \bibinfo{author}{\bibfnamefont{B.}~\bibnamefont{Hof}}, \bibnamefont{and}
  \bibinfo{author}{\bibfnamefont{J.}~\bibnamefont{Westerweel}},
  \bibinfo{journal}{Annu. Rev. of Fluid Mechanics}
  \textbf{\bibinfo{volume}{39}}, \bibinfo{pages}{447} (\bibinfo{year}{2007}).

\bibitem[{\citenamefont{Morris}(2009)}]{mor_ra09}
\bibinfo{author}{\bibfnamefont{J.}~\bibnamefont{Morris}},
  \bibinfo{journal}{Rheol. acta} \textbf{\bibinfo{volume}{48}},
  \bibinfo{pages}{909} (\bibinfo{year}{2009}).

\bibitem[{\citenamefont{Matas et~al.}(2003)\citenamefont{Matas, Morris, and
  Guazzelli}}]{Matas03}
\bibinfo{author}{\bibfnamefont{J.}~\bibnamefont{Matas}},
  \bibinfo{author}{\bibfnamefont{J.}~\bibnamefont{Morris}}, \bibnamefont{and}
  \bibinfo{author}{\bibfnamefont{{\'E}.}~\bibnamefont{Guazzelli}},
  \bibinfo{journal}{Physical Review Letter} \textbf{\bibinfo{volume}{90}}
  (\bibinfo{year}{2003}).

\bibitem[{\citenamefont{Yu et~al.}(2013)\citenamefont{Yu, Wu, Shao, and
  Lin}}]{Yu13}
\bibinfo{author}{\bibfnamefont{Z.}~\bibnamefont{Yu}},
  \bibinfo{author}{\bibfnamefont{T.}~\bibnamefont{Wu}},
  \bibinfo{author}{\bibfnamefont{X.}~\bibnamefont{Shao}}, \bibnamefont{and}
  \bibinfo{author}{\bibfnamefont{J.}~\bibnamefont{Lin}},
  \bibinfo{journal}{Physics Of Fluids} \textbf{\bibinfo{volume}{25}}
  (\bibinfo{year}{2013}).

\bibitem[{\citenamefont{Pope}(2000)}]{Pope00}
\bibinfo{author}{\bibfnamefont{S.}~\bibnamefont{Pope}},
  \emph{\bibinfo{title}{Turbulent flows}} (\bibinfo{publisher}{Cambridge
  University Press}, \bibinfo{year}{2000}).

\bibitem[{\citenamefont{Breugem}(2012)}]{Breugem12}
\bibinfo{author}{\bibfnamefont{W.-P.} \bibnamefont{Breugem}},
  \bibinfo{journal}{Journal of Computational Physics}
  \textbf{\bibinfo{volume}{231}}, \bibinfo{pages}{4469} (\bibinfo{year}{2012}).

\bibitem[{\citenamefont{Lambert et~al.}(2013)\citenamefont{Lambert, Picano,
  Breugem, and Brandt}}]{Lambert13}
\bibinfo{author}{\bibfnamefont{R.}~\bibnamefont{Lambert}},
  \bibinfo{author}{\bibfnamefont{F.}~\bibnamefont{Picano}},
  \bibinfo{author}{\bibfnamefont{W.~P.} \bibnamefont{Breugem}},
  \bibnamefont{and} \bibinfo{author}{\bibfnamefont{L.}~\bibnamefont{Brandt}},
  \bibinfo{journal}{J. Fluid Mech.} \textbf{\bibinfo{volume}{733}},
  \bibinfo{pages}{528} (\bibinfo{year}{2013}).

\bibitem[{\citenamefont{Picano et~al.}(2013)\citenamefont{Picano, Breugem,
  Mitra, and Brandt}}]{Picano13}
\bibinfo{author}{\bibfnamefont{F.}~\bibnamefont{Picano}},
  \bibinfo{author}{\bibfnamefont{W.-P.} \bibnamefont{Breugem}},
  \bibinfo{author}{\bibfnamefont{D.}~\bibnamefont{Mitra}}, \bibnamefont{and}
  \bibinfo{author}{\bibfnamefont{L.}~\bibnamefont{Brandt}},
  \bibinfo{journal}{Physical Review Letter} \textbf{\bibinfo{volume}{111}}
  (\bibinfo{year}{2013}).

\bibitem[{\citenamefont{Henningson and J.}(1991)}]{Henningson91}
\bibinfo{author}{\bibfnamefont{D.}~\bibnamefont{Henningson}} \bibnamefont{and}
  \bibinfo{author}{\bibfnamefont{K.}~\bibnamefont{J.}}, \bibinfo{journal}{J.
  Fluid Mech.} \textbf{\bibinfo{volume}{228}}, \bibinfo{pages}{183}
  (\bibinfo{year}{1991}).

\bibitem[{\citenamefont{Stickel and Powell}(2005)}]{stipow_arfm05}
\bibinfo{author}{\bibfnamefont{J.}~\bibnamefont{Stickel}} \bibnamefont{and}
  \bibinfo{author}{\bibfnamefont{R.}~\bibnamefont{Powell}},
  \bibinfo{journal}{Annu. Rev. Fluid Mech.} \textbf{\bibinfo{volume}{37}},
  \bibinfo{pages}{129} (\bibinfo{year}{2005}).

\bibitem[{\citenamefont{Yeo and Maxey}(2011)}]{yeomax_jfm11}
\bibinfo{author}{\bibfnamefont{K.}~\bibnamefont{Yeo}} \bibnamefont{and}
  \bibinfo{author}{\bibfnamefont{M.}~\bibnamefont{Maxey}},
  \bibinfo{journal}{Journal of Fluid Mechanics} \textbf{\bibinfo{volume}{682}},
  \bibinfo{pages}{491} (\bibinfo{year}{2011}).

\bibitem[{\citenamefont{Zhang and Prosperetti}(2010)}]{Zhang10}
\bibinfo{author}{\bibfnamefont{Q.}~\bibnamefont{Zhang}} \bibnamefont{and}
  \bibinfo{author}{\bibfnamefont{A.}~\bibnamefont{Prosperetti}},
  \bibinfo{journal}{Physics of Fluid} \textbf{\bibinfo{volume}{22}}
  (\bibinfo{year}{2010}).

\bibitem[{\citenamefont{Batchelor}(1970)}]{Batchelor70}
\bibinfo{author}{\bibfnamefont{G.~K.} \bibnamefont{Batchelor}},
  \bibinfo{journal}{J. Fluid Mech.} \textbf{\bibinfo{volume}{41}},
  \bibinfo{pages}{545} (\bibinfo{year}{1970}).

\bibitem[{\citenamefont{Brown and Jaeger}(2009)}]{brojae_prl09}
\bibinfo{author}{\bibfnamefont{E.}~\bibnamefont{Brown}} \bibnamefont{and}
  \bibinfo{author}{\bibfnamefont{H.}~\bibnamefont{Jaeger}},
  \bibinfo{journal}{Physical review letters} \textbf{\bibinfo{volume}{103}},
  \bibinfo{pages}{86001} (\bibinfo{year}{2009}).

\bibitem[{\citenamefont{Fall et~al.}(2010)\citenamefont{Fall, Lema{\^\i}tre,
  Bertrand, Bonn, and Ovarlez}}]{Fall10}
\bibinfo{author}{\bibfnamefont{A.}~\bibnamefont{Fall}},
  \bibinfo{author}{\bibfnamefont{A.}~\bibnamefont{Lema{\^\i}tre}},
  \bibinfo{author}{\bibfnamefont{F.}~\bibnamefont{Bertrand}},
  \bibinfo{author}{\bibfnamefont{D.}~\bibnamefont{Bonn}}, \bibnamefont{and}
  \bibinfo{author}{\bibfnamefont{G.}~\bibnamefont{Ovarlez}},
  \bibinfo{journal}{Phys. Rev. Lett.} \textbf{\bibinfo{volume}{105}}
  (\bibinfo{year}{2010}).

\bibitem[{\citenamefont{Bagnold}(1954)}]{bagnold}
\bibinfo{author}{\bibfnamefont{R.}~\bibnamefont{Bagnold}},
  \bibinfo{journal}{Proc. Royal Soc. London. Series A. Math. and Phys. Sci.}
  \textbf{\bibinfo{volume}{225}}, \bibinfo{pages}{49} (\bibinfo{year}{1954}).

\bibitem[{\citenamefont{Kulkarni and Morris}(2008)}]{kulmor_pof08}
\bibinfo{author}{\bibfnamefont{P.}~\bibnamefont{Kulkarni}} \bibnamefont{and}
  \bibinfo{author}{\bibfnamefont{J.}~\bibnamefont{Morris}},
  \bibinfo{journal}{Phys. Fluids} \textbf{\bibinfo{volume}{20}},
  \bibinfo{pages}{040602} (\bibinfo{year}{2008}).

\end{thebibliography}
\end{document}